\documentclass[sigconf, authorversion]{acmart}

\usepackage{todonotes}

\newcommand{\eg}{e.g.,~}
\newcommand{\ie}{i.e.,~}

\AtBeginDocument{%
  \providecommand\BibTeX{{%
    \normalfont B\kern-0.5em{\scshape i\kern-0.25em b}\kern-0.8em\TeX}}}

\usepackage{adjustbox}
\usepackage{listings}
\usepackage{color, colortbl}
\usepackage{multirow}
\usepackage{subcaption}

\usepackage{makecell}
\usepackage{csquotes}
\renewcommand{\mkbegdispquote}[2]{\itshape}

\newcommand{\boxit}[1]{\vspace{0.3cm}
\noindent
\begin{center}
\fbox{
\begin{minipage}{0.9\columnwidth}
\emph{#1} 
\end{minipage}
}
\end{center}
}

\lstdefinelanguage{Java}{
  tabsize=4
}[keywords,comments,strings]

\definecolor{source}{gray}{0.95}
\definecolor{highlight}{gray}{0.9}

\lstnewenvironment{codesnippet}{%
	\lstset{%
		frame=single,
		framerule=0pt,
		mathescape=false
	}
}{}

\copyrightyear{2020}
\acmYear{2020}
\setcopyright{acmcopyright}\acmConference[ESEM '20]{ESEM '20: ACM / IEEE International Symposium on Empirical Software Engineering and Measurement (ESEM)}{October 8--9, 2020}{Bari, Italy}
\acmBooktitle{ESEM '20: ACM / IEEE International Symposium on Empirical Software Engineering and Measurement (ESEM) (ESEM '20), October 8--9, 2020, Bari, Italy}
\acmPrice{15.00}
\acmDOI{10.1145/3382494.3410687}
\acmISBN{978-1-4503-7580-1/20/10}

\makeatletter                  
\def\mdseries@tt{m}      
\makeatother                   

\begin{document}

\title{Why Research on Test-Driven Development is Inconclusive?}

\author{Mohammad Ghafari}
\affiliation{\institution{University of Bern}}
\email{mohammad.ghafari@inf.unibe.ch}
\orcid{0000-0002-1986-9668}

\author{Timm Gross}
\affiliation{\institution{University of Bern}}
\email{timm.gross@id.unibe.ch}

\author{Davide Fucci}
\affiliation{\institution{Blekinge Institute of Technology}}
\email{davide.fucci@bth.se}

\author{Michael Felderer}
\affiliation{\institution{University of Innsbruck}}
\email{michael.felderer@uibk.ac.at}
\orcid{0000-0003-3818-4442}

\renewcommand{\shortauthors}{Ghafari, et al.}

\begin{abstract}

[Background] Recent investigations into the effects of Test-Driven Development (TDD) have been contradictory and inconclusive. This hinders development teams to use research results as the basis for deciding whether and how to apply TDD.
[Aim] To support researchers when designing a new study and to increase the applicability of TDD research in the decision-making process in industrial context, we aim at identifying the reasons behind the inconclusive research results in TDD.
[Method] We studied the state of the art in TDD research published in top venues in the past decade, and analyzed the way these studies were set up.
[Results] We identified five categories of factors that directly impact the outcome of studies on TDD.
[Conclusions] This work can help researchers to conduct more reliable studies, and inform practitioners of risks they need to consider when consulting research on TDD.

\end{abstract}

%


\maketitle

%
%

\section{Introduction}

Test-driven development (TDD) is a development technique---initially proposed twenty years ago~\cite{Beck:1999:EPE:318762}---in which failing tests are written before any code is added or changed.
This technique emphasizes small iterations and interleaved refactoring~\cite{madeyski2010test-driven}.

In the scientific literature, experts usually emphasize the positive effects of TDD~\cite{Shull2010, Buchan2011, Scanniello2016}.
This technique has become an integral part of the software engineering curriculum in universities~\cite{Kazerouni2019}.
When looking at the discourse around TDD in the grey literature, such as practitioners' blog posts or discussions, it becomes apparent that TDD has attracted great attention from practitioners---for instance, the ``TDD'' tag on Stack Overflow has 4.7k watchers.

The motivation for this work is to provide software companies a road map for the introduction of TDD in their policies based on the current state of research.
However, before that can happen, practitioners need to be made aware of the TDD research results, which are often inconclusive and oftentimes contradictory~\cite{Karac2018}. 

Although it is often claimed that TDD improves code quality (\eg results in fewer bugs and defects), one of the largest systematic studies in this domain~\cite{Munir2014a} shows that improvement in some studies is not significant, and that the claimed code quality gains are much more pronounced in ``low-rigor'' and ``low-relevance'' studies~\cite{Ivarsson:2011}.
Research has also studied the impact of TDD on the productivity of software developers---\eg in terms of generation of new code and effort required to fix bugs.
Some studies, for example~\citet{Kollanus2010}, claim that quality is increased at the price of degraded productivity; whereas some others, such as~\citet{Bissi2016}, argue that existing studies are inconclusive as, for example, experiments in an academic context are different from an industrial context.

These contradictions make it impossible to categorically provide evidence on the usefulness and effectiveness of TDD.
Therefore, in this paper, we focus on identifying major factors that render findings in this field inconclusive and hinder the applicability of TDD research in the decision-making process in industrial context.
Consequently, we answer the following research question: ``\emph{What factors can contribute to inconclusive research results on TDD?}''

To answer our research question, we studied, from the lens of a practitioner, the state of the art in TDD research.
We investigated contradictory results in this domain by studying secondary studies that organize the large body of research in the field.
We then focused on primary studies published in top journals and conferences in the past decade. We compared several studies that investigated similar phenomena (\eg internal or external code quality) to identify factors that may contribute to inconclusive results in TDD.

%

We identified five categories of factors concerning how studies are set up that contribute to this problem.
These categories are TDD definition, participants, task, type of project, and comparison.
We found that the exact definition of TDD that a study follows is not always clear;
the participants of the studies are often newcomers to this technique;
experiments mainly focus on code generation in greenfield projects, and the opportunities to adopt TDD in an existing codebase is not investigated;
the baseline practice against which TDD is compared should be agile;
and finally, exploration of the long-term benefits and drawbacks of TDD has not received enough attention in the literature.

\newpage 

In summary, 
this paper is the first to survey factors related to inconclusive results in TDD research.
We believe it has important implications for both researchers and practitioners.
It paves the way for researchers to conduct more reliable studies on TDD, and alert practitioners of important factors that they should consider when seeking advise from research in this area.

The rest of this paper is structured as follows.
In Section~\ref{section:methodology}, we explain the methodology we followed to conduct this study. 
In Section~\ref{section:results} we present our findings.
We discuss the implications of this research for practitioners and researchers in Section~\ref{section:discussion}.
In Section~\ref{section:threats}, we discuss the threats to validity of this work, and we conclude the paper in Section~\ref{section:conclusion}.
%
%

\section{Methodology}
\label{section:methodology}

We conducted a literature study to compile a list of factors that are responsible for diverging research results and hinder the applicability of TDD research in practice.
We were interested in threats that have an explicit impact on TDD and excluded those that, for instance, are inherent to the type of a study such as hypothesis guessing or evaluation apprehension in controlled experiments.

\begin{figure*}
  \centering
  \includegraphics[scale=0.7]{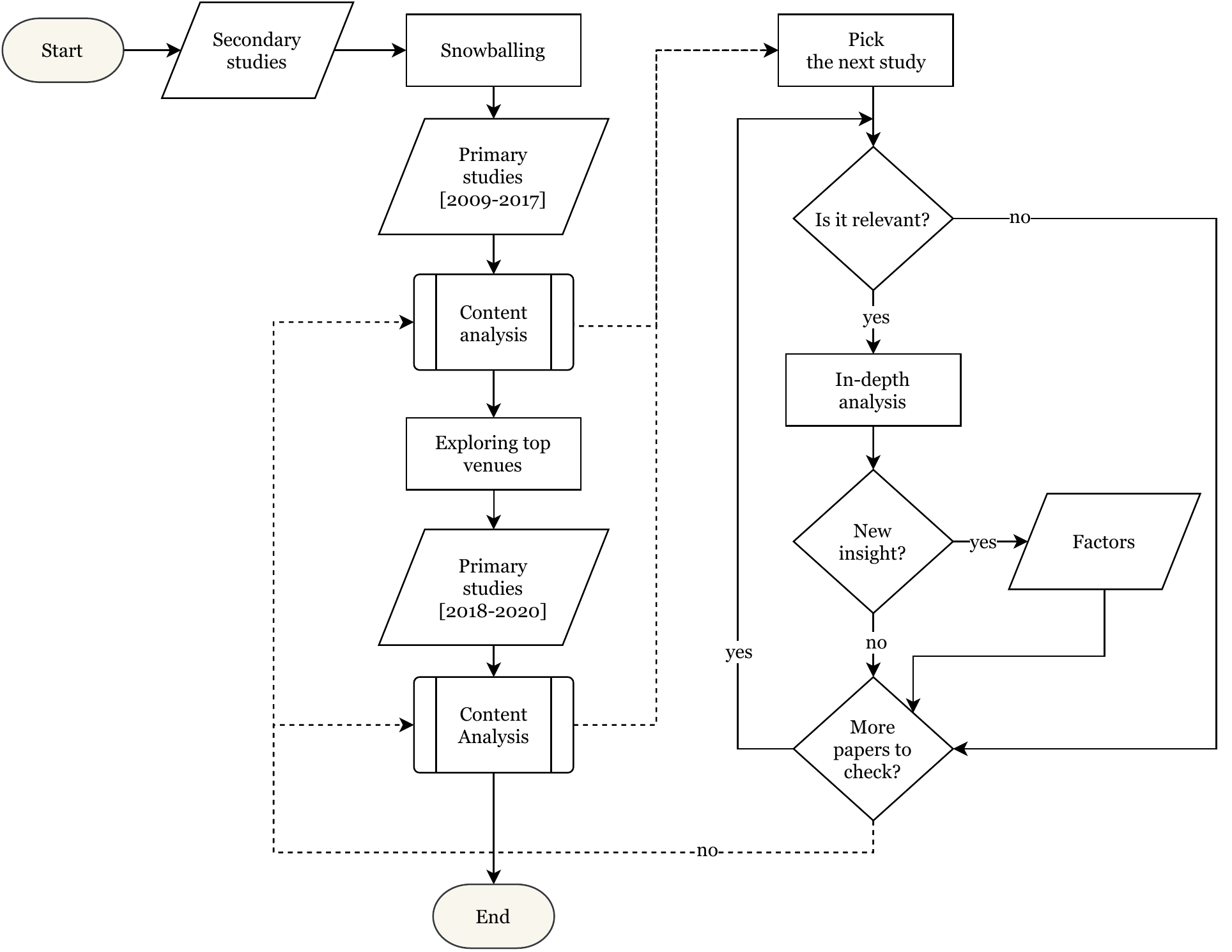}
  \caption{The methodology of our literature review.}
  \label{fig:steps}
\end{figure*}

We followed three main steps.
Firstly, we studied literature reviews that concern TDD to acquaint ourselves with the state of research in this area, and to build an overview of the diverging results.
We followed backward snowballing to obtain a list of primary studies from these literature reviews that were published from 2009 to 2017.
Secondly, we analyzed these primary studies to identify reasons for inconclusive research into TDD.
Thirdly, we went through the proceedings of several top journals/conferences, and collected papers published after the latest review study (\ie from 2018 to April 2020) to capture the most recent work in the field.
In the following we discuss these steps in detail as shown in Figure~\ref{fig:steps}.

In the \textbf{first step}, we looked at secondary studies on TDD.
We mainly based our work on nine secondary studies reported in a recent meta literature study~\cite{Karac2018}.
We used these secondary studies (see Table \ref{tab:litreviews}) to get an overview of the state of research on TDD, and to acquaint ourselves with the diverging results discussed in previous work.

\begin{table*}
\caption{The secondary studies we analyzed in the first step}
\label{tab:litreviews}
\begin{tabular}{ll}
\multicolumn{1}{c}{\textbf{Authors}} & \multicolumn{1}{c}{\textbf{Title}}\\ \hline

\citet{Karac2018} &
What Do We (Really) Know about Test-Driven Development?
\\ \hline

\citet{Bissi2016} &
The effects of test driven development on internal quality, external quality and productivity: A systematic review
\\ \hline

\citet{Munir2014a} & 
Considering rigor and relevance when evaluating test driven development: A systematic review
\\ \hline

\citet{Rafique2013} & 
The Effects of Test-Driven Development on External Quality and Productivity: A Meta-analysis 
\\ \hline

\citet{Causevic2011} &
Factors Limiting Industrial Adoption of Test Driven Development: A Systematic Review
\\ \hline

\citet{Shull2010} & 
What Do We Know about Test-Driven Development? 
\\ \hline

\citet{Turhan2010} & 
How Effective is Test-Driven Development? 
\\ \hline

\citet{Kollanus2010} & 
Test-Driven Development - Still a Promising Approach?
\\ \hline

\citet{Siniaalto2006} & 
Test driven development: empirical body of evidence
\\ 

\end{tabular}
\end{table*}

From these literature reviews we followed backward snowballing to identify potential primary studies to include in this analysis.
We did not select studies published earlier than 2009.
The decision to focus on publications in the past decade was mainly due to our limited resource that we prioritized on more recent body of knowledge in the field.

We then started with the \textbf{second step}, the iterative identification and refinement of the factors that contribute to diverging outcomes in research on TDD.
In order to achieve this, we had to reason about explicit and implicit threats to validity of TDD studies. 
However, the way each study was reported varied.
We, the first two authors of this paper, read each study thoroughly, filled in a data extraction form, and resolved any conflict by discussion.
We picked one primary study and analyzed its goals, setup, execution, findings, and threats to validity.
We compared studies that investigated similar goals, for instance, assessing the impact of TDD on internal or external code quality.
We then used the results of our analysis to firstly, refine our list categories of factors, either by adding a new category or by sharpening an existing category, and to secondly provide examples of the existing categories.
Next, we picked another primary study and repeated this process.

The selection process for the next paper chosen to be analysed was based on two criteria.
First, we preferred studies that were cited multiple times and for which the abstract sounded relevant (\eg it explains a comparative study or measures the impact of TDD).
Secondly, we tried to keep a balance between the different types of studies such as experiments, case studies, and surveys.

To determine when to stop the iteration, we used a criterion of saturation --- \ie we stopped adding new primary studies once the inclusion of a new one did not reveal a new threat nor provided any additional information regarding one of the identified categories of factors.
Table~\ref{tab:primary-1} lists ten carefully selected examples of primary studies that we analyzed in this step.

\begin{table*}
\caption{Examples of the primary studies collected in the second step}
\label{tab:primary-1}
\begin{tabular}{ll}
\multicolumn{1}{c}{\textbf{Authors}} & \multicolumn{1}{c}{\textbf{Title}}\\ \hline

\citet{Pancur2011} & Impact of test-driven development on productivity, code and tests: A controlled experiment \\ \hline
\citet{Fucci2017} & A Dissection of the Test-Driven Development Process: Does It Really Matter to Test-First or to Test-Last? \\ \hline

\citet{Batic2011} & The effectiveness of test-driven development : an industrial case study \\ \hline
\citet{Fucci2013} & A Replicated Experiment on the Effectiveness of Test-first Development\\\hline

\citet{Thomson2009} & What Makes Testing Work: Nine Case Studies of Software Development Teams \\ \hline
\citet{Romano2017} & Findings from a multi-method study on test-driven development \\ \hline
\citet{Buchan2011} & Causal Factors, Benefits and Challenges of Test-Driven Development: Practitioner Perceptions\\ \hline
\citet{Scanniello2016} & Students' and Professionals' Perceptions of Test-driven Development: A Focus Group Study\\ \hline
\citet{Beller2017} & Developer Testing in The IDE: Patterns, Beliefs, And Behavior\\ \hline

\citet{Bannerman2011} & A multiple comparative study of test-with development product changes and their effects on team speed\\& and product quality\\

\end{tabular}
\end{table*}

In the \textbf{third step}, we reflected on recent studies in the field.
We browsed the proceedings of top-tier conferences and issues of journals from 2018 to April 2020 to include papers published after the latest TDD review study.\footnote{
We mainly selected top relevant journals from the ISI listed journals, and consulted the core conference ranking to identify relevant venues with at least A ranking.}
We searched for the terms ``TDD'', ``test driven'', ``test-driven'', ``test first'', and ``test-first'' in several top-tier journals/conferences.
Particularly, we looked at six Journals (IEEE Transactions on Software Engineering, Empirical Software Engineering, Software Testing, Verification, and Reliability Journal, Journal of Systems and Software, Information and Software Technology, and Journal of Software: Evolution and Process);
the proceedings of eight Software Engineering Conferences (International Conference on Software Engineering, International Conference on Automated Software Engineering, Joint European Software Engineering Conference and Symposium on the Foundations of Software Engineering, International Conference on Software Analysis, Evolution and Reengineering, International Conference on Software Maintenance and Evolution, International Symposium on Empirical Software Engineering and Measurement, International Conference on Evaluation and Assessment in Software Engineering, and International Conference on Mining Software Repositories); 
two top testing conferences (International Conference on Software Testing, Verification and Validation, International Symposium on Software Testing and Analysis);
and three Software Process Conferences (International Conference on Agile Software Development, International Conference on Software and Systems Process, International Conference on Product-Focused Software Process Improvement).
This process resulted in only ten new papers listed in table~\ref{tab:primary-2}.
We studied each paper in depth, similarly to the primary studies in the previous step, to check whether we can obtain a new insight.

\begin{table*}
\caption{The primary studies collected in the third step}
\label{tab:primary-2}
\begin{tabular}{ll}
\multicolumn{1}{c}{\textbf{Authors}} & \multicolumn{1}{c}{\textbf{Title}}\\ \hline

\citet{Karac2019} & A Controlled Experiment with Novice Developers on the Impact of Task Description Granularity on\\&Software Quality in Test-Driven Development \\ \hline
\citet{Tosun2019} & Investigating the Impact of Development Task on External Quality in Test-Driven Development: An\\& Industry Experiment\\ \hline

\citet{Borle2018} & Analyzing the effects of test driven development in GitHub\\ \hline
\citet{Fucci2018} & A longitudinal cohort study on the retainment of test-driven development \\ \hline
\citet{Kazerouni2019} & Assessing Incremental Testing Practices and Their Impact on Project Outcomes \\ \hline
\citet{Santos2018} & Improving Development Practices through Experimentation : an Industrial TDD Case\\ \hline

\citet{Tosun2018} & On the Effectiveness of Unit Tests in Test-driven Development\\\hline

\citet{B2018}& Does the Performance of TDD Hold Across Software Companies and Premises? A Group of Industrial Experiments on TDD \\ \hline

\citet{Romano:2019} & An Empirical Assessment on Affective Reactions of Novice Developers When Applying Test-Driven Development\\ \hline

\citet{Sundelin:2018} & Test-Driving FinTech Product Development: An Experience Report \\

\end{tabular}
\end{table*}

%
%

\section{Results}
\label{section:results}

\begin{figure}[htb]
  \includegraphics[width=\columnwidth]{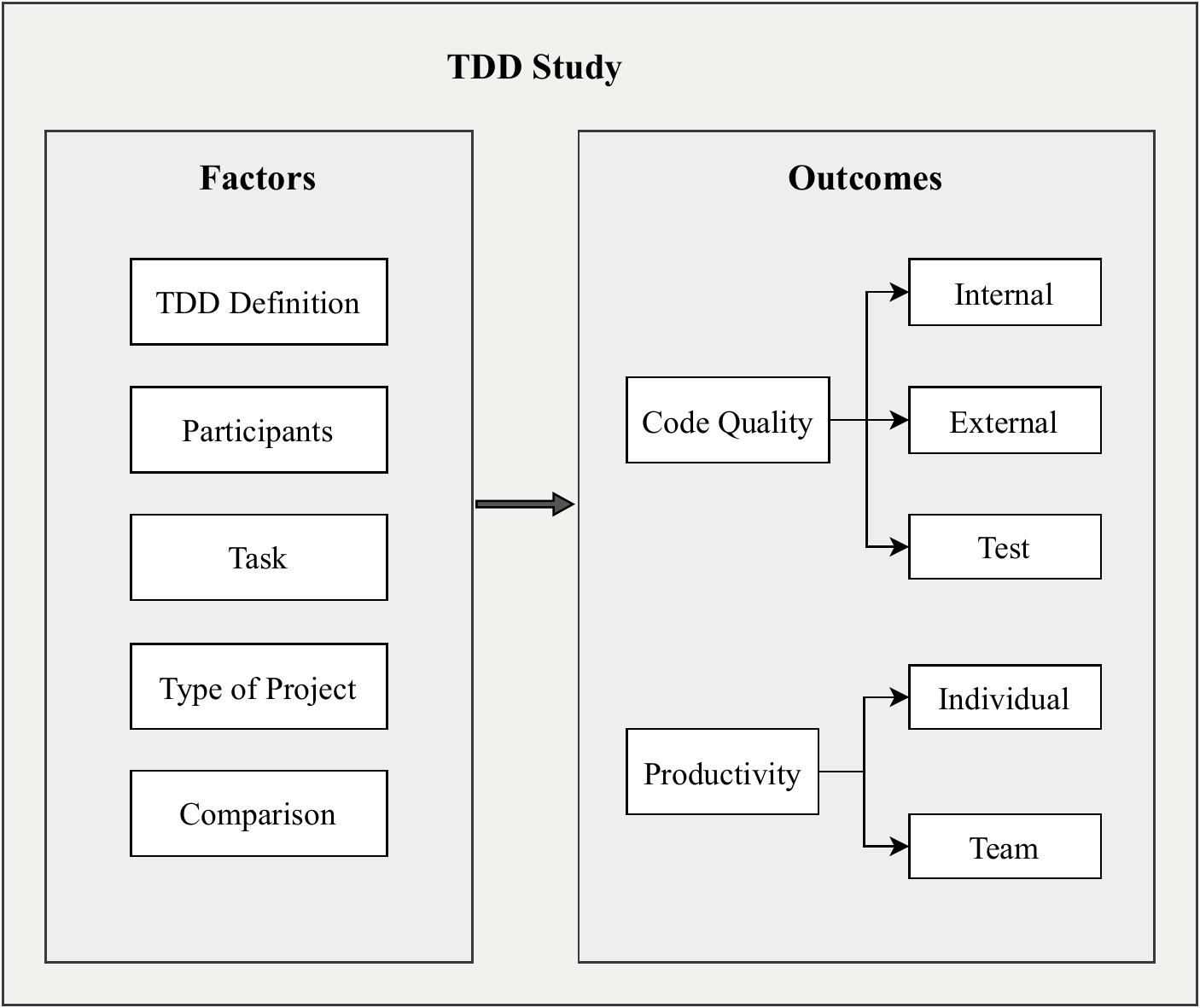}
  \caption{Factors contributing to the inconclusive outcomes in research on TDD.}
  \label{fig:overview}
\end{figure}

There have been many investigations into understanding the outcome of TDD in software development.
Nevertheless, the understanding of the different outcomes of TDD is still inconclusive due to several reasons lying in the way previous studies were set up.
In this section we discuss these outcomes and factors responsible for contradictory understanding, which is summarized in Figure~\ref{fig:overview}.

\subsection{Outcomes}

In general, TDD promises to improve developer productivity and three dimensions of code quality, namely internal and external code quality as well as test quality~\cite{Beck2002}.
External code quality is usually relevant for the users and is measured in terms of how well the code covers and implements the requirements or user stories. 
Internal code quality is only relevant for developers and describes how well the code is structured, how complex it is to understand or how maintainable it is.

\begin{table*}
\caption{Measurement of internal code and test quality}
\label{tab:intqual}
\begin{tabular}{p{3cm}|p{8cm}}
Complexity & \citet{Pancur2011}, \citet{Batic2011}, \citet{Bannerman2011}, \citet{Tosun2019}\\ \hline
Code coverage & \citet{Tosun2018}, \citet{Pancur2011}, \citet{Kazerouni2019}, \citet{Thomson2009}, \citet{Borle2018}, \citet{Bannerman2011}\\ \hline
Mutation score & \citet{Tosun2018}, \citet{Pancur2011}\\ \hline
None & \citet{Fucci2017}, \citet{Fucci2018}, \citet{Fucci2013}, \citet{Santos2018}, \citet{Beller2017}, \citet{Karac2019}\\ 
\end{tabular}
\end{table*}

There are several ways to measure internal (code and test) quality (see Table~\ref{tab:intqual}).
For instance,~\citet{Shull2010} reviewed studies that measured code quality in terms of metrics such as coupling and cohesion, complexity, and density.
They reported mixed results with some papers measuring better and others measuring worse internal code quality.

In terms of test quality, research has explored the quality of tests by measuring mutation scores (\ie the bug detection ability of the tests) and code coverage (\ie the degree to which the source code of a program is executed when a test suite runs).
For example,~\citet{Tosun2018} conducted an experiment with 24 professionals and found that unit-test cases developed in TDD have a higher mutation score and branch coverage, but less method coverage than those developed in ITL.
Their findings contradicts earlier findings that were mostly conducted with students~\cite{Madeyski2010}.

In terms of external quality and developer productivity, previous research has mostly investigated new code generation (\eg accepted user stories and time to implement them).
For instance,
\linebreak
~\citet{Marchenko2009} interviewed eight participants who used TDD at Nokia-Siemens Network for three years.
The participants stated that the team confidence with the code base is improved, which is associated with improved productivity.
~\citet{Fucci2018} conducted an experiment with students over a period of five months and showed that adoption of TDD only results in writing more tests; otherwise it has neither statistically significant effect on the external quality of software products nor on the developers’ productivity.

We noted that TDD research has looked at bugs and code maintainability as static indicators for external and internal quality, respectively. 
However, in practice, their costs would be manifested to the full extent only once software is in use.
Especially, we rarely found studies on the maintainability of tests and their co-evolution with production code.
One reason might be that many people do not consider TDD as a testing technique per se, but as a design technique~\cite{Beck2002}.
However,~\citet{Sundelin:2018} studied a financial software under development for eight years, and found that the size of tests grows much faster than of production code.
Therefore, it is necessary to clean, refactor, and prioritize tests to manage this grows.

\boxit{
Research often deals with short-term impact of TDD rather than its long-term benefits and drawbacks, which manifest themselves once the software is in use.
This is especially the case for quality of test suites.
}


\subsection{Factors}

We identified five categories of factors, namely TDD definition, participants, task, type of project, and comparison that influence the outcome of TDD research.
In the following, we present these categories in detail.

\subsubsection{TDD definition}

The steps defining TDD and how strictly they are followed is very important for a study.
There are two common TDD styles: one is classical TDD, where there is almost no design upfront and developers just drive the entire implementation from the tests; and the other one is where developers know the design before developing~\cite{kahneman:2015}.
In effect, developers often adopt a combination of these styles depending on the problem domain.
However, we noted that a commonly shared definition of TDD is missing.
What TDD means is mostly boiled down to writing tests prior to production code, and its other characteristics have not received similar attention. 
For example, some studies measure refactoring explicitly and even use it to assess how much participants adhere to TDD, while others are not concerned with refactoring, even though it is supposed to be a key part of TDD~\cite{Beck2002}.

There are a few recent studies that investigated how testing is actually done ``in the wild''.
~\citet{Beller2017} observed the work of 2,443 software developers over 2.5 years and discovered that developers who claim to do TDD, neither follow it strictly nor for all their modifications.
They found that only 2.2\% of sessions with test executions contain strict TDD patterns.
~\citet{Borle2018} showed that TDD is practiced in only 0.8\% of the 256,572 investigated public GitHub projects which contain test files.

\boxit{There is a variety of TDD definitions. Its exact meaning, the underlying assumptions, and how strictly one follows it are not well-explained in previous studies.}


\subsubsection{Participants selection}

Studies who recruit their participants from companies tend to have fewer participants than studies done with students. 
One can see that from Table~\ref{tab:participants}, which shows the numbers of participants in industrial and academic studies.
In particular, studies with professionals usually have a maximum of 20 participants, whereas studies with students have in several cases 40+ participants.

\begin{table*}
\caption{Population of participants in studies with students and professionals}
\label{tab:participants}
\begin{tabular}{p{2cm}|p{4cm}|p{4cm}|p{4cm}}
& \multicolumn{1}{c}{\textbf{Less than 20 participants} }   & \multicolumn{1}{c}{\textbf{21-40 participants}} & \multicolumn{1}{c}{\textbf{More than 40 participants}} \\\hline
Industrial & \citet{Romano2017}, \citet{Buchan2011}, \citet{Scanniello2016}, \citet{Santos2018}, \citet{Tosun2019} & \citet{Tosun2018}, \citet{Batic2011}, \citet{Fucci2017}&\\\hline
Academic & \citet{Romano2017}, \citet{Scanniello2016} & \citet{Thomson2009} & \citet{Pancur2011}, \citet{Kazerouni2019}, \citet{Fucci2013}, \citet{Karac2019} \\
\end{tabular}
\label{participants}
\end{table*}

We observed that experiments are mostly conducted as part of exercises in a one-semester course with students, whereas in industry they are often part of an intensive course with professional participants lasting a couple of days (see Table~\ref{tab:experience}).
Nevertheless, anecdotal~\cite{Shull2010} as well as empirical evidence~\cite{Scanniello2016} suggest that when introducing TDD to developers, the benefits manifest themselves only after an initial investment and a ramp-up time.
We noted that studies with participants who are proficient in TDD prior to the start of experiments, for example~\cite{Buchan2011}, are in the minority.
We even observed studies, for example~\cite{Tosun2019}, where participants were asked to follow TDD right after only a short introduction.

The fact that both practitioners and students have quite similar TDD experience (\ie they have undergone very little training in TDD) does not necessarily imply that when practicing TDD the outcomes of the two subject groups are also similar.
Professionals' competencies, for instance to develop tests and design software, may influence their performance when practicing TDD.
For instance,
\citet{B2018} conducted four industrial experiments in two different companies, and reported that the larger the experience with unit testing and testing tools, the better developers perform in terms of external quality in ITL than in TDD.
\citet{Latorre:2014} found that in unit test-driven development, junior developers are not able to discover the best design, and this translates into a performance penalty since they need to revise their design choices more frequently than skilled developers.
\citet{Romano:2019} investigated the affective reactions of novice developers to the development approach and reported that novices seem to like a non-TDD development approach more than TDD, and that the testing phase makes developers using TDD less happy.
~\citet{Suleman2017} conducted an early pilot study with students who experienced TDD in an introductory programming course.
They found that students do not necessarily experience the immediate benefits of TDD, and that this TDD is perceived to be more of a hindrance than a help to them.

\boxit{
Studies participants (i.e., students and professionals) have little prior TDD experience, ranging generally from a couple of days to a couple of months. 
}

\begin{table*}
\caption{TDD experience}
\label{tab:experience}
\begin{tabular}{p{3cm}|p{8cm}}
<1 week & \citet{Tosun2018}, \citet{Fucci2017}, \citet{Thomson2009}, \citet{Santos2018}, \citet{Tosun2019}\\\hline
1 week - 0.5 years & \citet{Fucci2018}, \citet{Kazerouni2019}, \citet{Romano2017}, \citet{Scanniello2016}, \citet{Batic2011},\\& \citet{Fucci2013}, \citet{Karac2019}\\\hline
0.5 years - 1 year & \citet{Pancur2011}\\\hline
more & \citet{Buchan2011}
\end{tabular}
\label{experience}
\end{table*}


\subsubsection{Task selection}

The number as well as the types of performed tasks are important.
Tasks that are synthetic are easily comparable, for example, in terms of complexity. 
Nevertheless, they do not resemble tasks assigned during the course of a real-world project.
We observed that most studies were concerned with one and up to four synthetic tasks, such as coding katas.
Table \ref{tab:tasksyn} shows which studies used what kind of tasks.
Surprisingly, synthetic tasks are dominant in experiments conducted in industrial settings.

\begin{table*}
\caption{Synthetic tasks vs. real-world tasks}
\label{tab:tasksyn}
\begin{tabular}{p{3cm}|p{8cm}}
Synthetic task & 
\citet{Romano2017}, \citet{Fucci2013}, \citet{Tosun2018}, \citet{Pancur2011}, \citet{Karac2019}, \citet{Tosun2019},\\& 
\citet{Fucci2017}, \citet{Santos2018},
\citet{Fucci2018}, \citet{Kazerouni2019}\\ \hline
Real task & \citet{Thomson2009}, \citet{Batic2011}\\ 
\end{tabular}
\end{table*}

\newpage

The granularity as well as the complexity of a task, \eg whether it is related to other parts of a software and whether developers are familiar with the task, may impact the TDD outcomes.
For instance,
~\cite{Karac2019} investigated the effect of task description granularity on the quality (functional correctness and completeness) of software developed in TDD by novice developers (precisely graduate students), and reported that more granular task descriptions significantly improve quality.
~\citet{Latorre:2014} showed that experienced developers who practice TDD for a short while become as effective in performing ``small programming tasks'' as compared to more traditional test-last development techniques.
However, many consider TDD as a design technique~\cite{Beck2002}, but how much design is involved in a small task is debatable.
Moreover, the suitability of TDD may differ not only for different tasks, but also for different parts in a software---\ie one might apply TDD to implement features in more critical parts of the code base and do not apply it for less critical parts.

Finally, previous literature is mostly concerned with code generation, and exploring how TDD performs during bug-fixing or large-scale refactoring has not received enough attention.
For instance,~\citet{Marchenko2009} interviewed a team of eight developers who adopted TDD at Nokia-Siemens Network for three years.
The team reported that TDD was not suitable for bug fixing, especially for bugs that are difficult to reproduce or for quick ``hacks'' due to the testing overhead.

\boxit{
Synthetic, non-real world tasks are dominant.
Research does not cover the variety of tasks to which TDD can be applied.
}


\subsubsection{Type of Project}

\begin{table*}
\caption{Green- vs. brownfield projects}
\label{tab:typefield}
\begin{tabular}{p{3cm}|p{8cm}}
Greenfield & \citet{Tosun2018}, \citet{Pancur2011}, \citet{Fucci2017}, \citet{Fucci2018}, \citet{Kazerouni2019}, \citet{Romano2017},\\& \citet{Thomson2009}, \citet{Batic2011}, \citet{Fucci2013}, \citet{Santos2018}, \citet{Karac2019}, \citet{Tosun2019}\\ \hline
Brownfield & \citet{Buchan2011}, \citet{Scanniello2016}
\end{tabular}
\end{table*}

In agile software development, developers are often involved in changing existing code, either during bug fixing or to implement changing requirements.
Therefore, whether the studies are concerned with projects developed from scratch (\ie greenfield), or with existing projects (\ie brownfield) plays a role.\footnote{Creating a new functionality in an existing project that is largely unrelated to the rest of the project is still a greenfield project.}
Brownfield projects are arguably closer to the daily work of a developer, and generalizing the results gathered from greenfield projects to brownfield projects may not be valid.
Nevertheless, brownfield projects are under-represented in existing research (see Table~\ref{tab:typefield}).

We believe that the application of TDD in an existing codebase depends on the availability of a rich test suite and the testability of a software --- \ie how difficult it is to develop and run tests~\cite{Ghafari:2019}.
In legacy systems that lack unit test cases, TDD may not be applicable as developers are deprived of the quick feedback from tests on changes.
However, understanding how TDD performs in brownfield projects that comprise regression test suites is a research opportunity that needs to be explored.

\boxit{
Research mostly focuses on greenfield projects rather than brownfield projects.
Accordingly, the opportunity to apply TDD in an existing codebase is unclear.
}


\subsubsection{Comparisons}

Factors that are actually responsible for the benefits of TDD vary.
For instance, research has shown that, when measuring quality, the degree of iteration of the process is more important than the order in which the test cases are written~\cite{Fucci2017}.
In a recent study, \citet{Karac2019} suggest that the success of TDD is correlated with the sub-division of a requirement into smaller tasks, leading to an increase in iterations.

Previous research has shown that a lot of the superiority of TDD in existing studies is the result of a comparison with a coarse-grained waterfall process\cite{Pancur2011}.
Nevertheless, TDD is an agile technique and should be compared with fine-grained iterative techniques, such as iterative test last (ITL), that share similar characteristics.
This means not only we do not know what exactly is responsible for the observed benefits of TDD, but also that the benefits we measure depend on what we compare TDD against.

\begin{table*}
\caption{What TDD is compared to}
\label{tab:comparisonstotdd}
\begin{tabular}{p{3cm}|p{8cm}}
Iterative test last & \citet{Tosun2018}, \citet{Pancur2011}, \citet{Kazerouni2019}, \citet{Fucci2017}, \citet{Santos2018}, \citet{Tosun2019}\\ \hline
Test last & \citet{Batic2011}, \citet{Fucci2013}, \citet{Bannerman2011}, \citet{Fucci2017}\\ \hline
Your way & \citet{Fucci2018}, \citet{Thomson2009}, \citet{Romano2017}, \citet{Santos2018}, \citet{Beller2017}, \citet{Buchan2011},\\& \citet{Scanniello2016}, \citet{Borle2018}\\ \hline
TDD & \citet{Karac2019}\\
\end{tabular}
\end{table*}

Table \ref{tab:comparisonstotdd} shows examples of what the analyzed studies compare TDD to.
``Test last'' (TL) describes that the tests are written after the production code without specifying when exactly.
``Iterative test last'' (ITL) is similar in that the tests are written after the production code is implemented, but it is supposed to have the same iterativeness as TDD.
This means in ITL a small code change is written and the tests are written immediately afterwards. 
The category ``Your way'' means that there is no guideline and developers should decide, if ever, when and how they write tests.
Finally, the category ``TDD'' compares TDD to itself in different settings.  
For instance, the performance impact the granularity of task description has on TDD~\cite{Karac2019}.

There may be more factors at play when comparing two techniques.
For instance, a recent work has shown that testing phase makes novice developers using TDD less happy~\cite{Romano:2019}.
In the same vein, students perceive TDD more of an obstacle than a help~\cite{Suleman2017}.
The affective reactions of developers may not have an immediate impact on the outcome of TDD, but exploring the consequences over the long run is necessary to draw fair conclusions.

\boxit{
The benefits of TDD may not be only due to writing tests first and, therefore, it should be compared to other Agile techniques.
}

%
%

\section{Discussion}
\label{section:discussion}

The promise of TDD is that it should lead to more testable and easier to modify code~\cite{Beck2002}.
This makes it appealing from an industrial perspective, as developers spend half of their time dealing with technical debt, debugging, and refactoring with an associated opportunity cost of 85\$ billion~\cite{Stripe.com2018}.
Nevertheless, the empirical evidence on TDD is contradictory, which hinders the adoption of this technique in practice.

\citet{Causevic2011} explored the reasons behind the limited industrial adoption of TDD, and identified seven factors, namely increased development time, insufficient TDD experience/knowledge, lack of upfront design, domain and tool specific issues, lack of developer skill in writing test cases, insufficient adherence to TDD protocol, and legacy code.
\citet{Munir2014a} investigated how the conclusions of existing research change when taking into account the relevance and rigor of studies in this field.
They found that studies with a high rigor and relevance scores show clear results for improvement in external quality at the price of degrading productivity.

We have built on previous work by exploring the latest state of the research in this domain. We identified factors that contribute to diverging results when studying TDD, and highlighted research opportunities that improve the applicability of research results for practitioners.
In particular,
we found that the exact definition of TDD that a study follows is not always clear;
the participants of the studies are often newcomers to this technique and experiments with TDD proficient participants are in a minority;
experiments mainly focus on code generation in greenfield projects, and the opportunities to adopt TDD in an existing codebase is not investigated;
the baseline practice against which TDD is compared should share similar agile characteristics;
and exploration of the long-term benefits and drawbacks of TDD, especially how to manage the large body of test cases generated in TDD, has not received enough attention in the literature.

This work has implications for both practitioners deciding on the adoption of TDD and researchers studying it.
We discuss these implications in the following.


\paragraph{Implications for practitioners}

We propose a list of known factors for practitioners to take into account when making a decision about TDD.
The factors are tuned for practitioners as their interest can be different from the one constituting the phenomena studied in research. 
For example, although a study may investigate the effect of TDD on maintainability (\ie an important aspect for a practitioner), it does so in a greenfield project (\ie irrelevant for the practitioners' everyday situation).
Therefore, the factors can be used as a support for practitioners navigating the (vast) scientific TDD literature and can be used to filter results interesting for their specific cases.

In general, industry practitioners are concerned that a low participation of professionals as subjects reduces the impact of software engineering research~\cite{FJW18}.
For practitioners, it is difficult to make a decision based on a group of students benefiting from TDD. 
Although CS graduates and entry-level developers are assumed to have similar skills~\cite{FJW18}, practitioners basing their decision to include TDD in their set of practices using the \emph{Participants} factor need to be aware that motivations between these two types of participants are different~\cite{FZB18}. 
Practitioners need to be also aware that designing experiments with students is vastly easier compared to professionals (\eg due to ease of recruitment).
Therefore, it is unwise to disregard potential insights gained from study with students.
Notably, the correct application of TDD requires training and practice~\cite{Kazerouni2019}, but the current investigations are manily based on the observation of practitioners (either professional or not) who often received a short crash course in TDD.
\citet{B2018} have shown that the larger the experience with unit testing and testing tools, the more developers outperform in ITL than in TDD.


\paragraph{Implications for researchers}

The factors presented in this study can serve as the basis for the development of guidelines on how to design TDD studies that result in converging results.
Similarly, researchers wanting to perform TDD studies---independently from their goal---need to prioritize the factors presented in this paper to be relevant for practice.

One factor we deem important for scientific investigation of TDD is \emph{Comparison}---\ie the baseline practice against which TDD is compared.
The IT landscape was different when the Agile methodologies, including TDD, were first proposed~\cite{Beck:1999:EPE:318762, Beck2002}. 
Not only the technologies, such as testing frameworks and automation infrastructure were not as mature as they are today, but also the development paradigms were mostly akin to the waterfall model, often without any explicit testing during development.
But now, 20 years later, it is necessary to re-evaluate what factors of TDD we study and what we compare TDD to.

We noted that research has mostly focused on short terms benefit (if any) of TDD, while it does not concentrate on how TDD impacts downstream activities in the software development life-cycle---\eg system testing~\cite{Off18}.
Similarly, understanding effects such as the actual maintenance costs that manifest themselves only when the software is in use has not received enough attention in research.
Especially, test suites could grow faster than production code in TDD~\cite{Sundelin:2018}, but we have not seen any study that concern managing tests.


\paragraph{Final remarks}

The major software testing venues do not seem to be interested in TDD---\eg no papers were published at the past two editions of ICST\footnote{International Conference on Software Testing}, ISSTA\footnote{International Symposium on Software Testing and Analysis}, ICSE\footnote{International Conference on Software Engineering}, and FSE\footnote{International Conference on the Foundations of Software Engineering} nor submitted to STVR\footnote{Software Testing, Verification, and Reliability Journal} between 2013 and 2020~\cite{Off18}.
We believe that addressing these factors is necessary for a \textit{renaissance} of TDD in the research community after the initial 15 years of inconclusive evidence.

It is noteworthy that the list of factors we presented in this paper, although grounded in the existing literature, is not exhaustive as several other factors apply specifically to industry. For instance, factors such as \textit{Agility} of a company~\cite{HDG06}, testing polices~\cite{HSF12}, and developers' work load have not received attention in research on TDD.
We believe that conducting general and convincing studies about TDD is hard, however, if TDD research is to be relevant for decision makers, more in-depth research is necessary to provide a fair account of problems in TDD experiments.

%
%

\section{Threats to Validity}
\label{section:threats}

We relied on several secondary studies to obtain a list of research on TDD which is as exhaustive as possible. 
We then manually browsed top and relevant journals/conferences to include recent papers.
However, there is always risk of omitting relevant papers when performing a literature study.
We mitigated the risk in two ways.
First, we clearly defined and discussed what primary studies fit the scope of our study, and conducted a pilot study to examine our decision criteria on whether or not to include a paper based on an iterative saturation approach.
Secondly, a random set of 15 excluded papers were examined independently by a second researcher to minimize the risk of missing important papers. 

The secondary studies used as a starting point in our process are Systematic Reviews and Meta-analyses which mainly aggregate evidence from quantitative investigations, such as controlled experiments.
Conversely, none of the secondary studies presented an aggregation of qualitative investigations, such as thematic or narrative synthesis~\cite{CDR15}.
Although this can result in a set of primary studies skewed towards one type of investigation, we made sure that each factor is reported in studies following both qualitative and quantitative research methodologies.

We sorted primary studies, published until 2017, according to number of citations.
We acknowledge that due to such a criterion, we may have failed to include more recent studies as they had less time to be cited. 
For more recent primary studies that we collected manually, published from 2018 to 2020, we included all the papers.

We had to understand, from the lens of practitioners, why research results on TDD are diverging and under which circumstances the results may not be generalizable to real-world context.
We treated papers as artifacts to be understood through qualitative literature analysis~\cite{flick2009introduction}, and tried to truthfully make connections between studies.
In order to mitigate the risk of missing or misinterpreting information from a study, we designed a data extraction form and discussed it together to develop a shared understanding.
We ran a pilot study with five randomly selected primary studies to make sure that we all agree on the extracted information.
Finally, through constant iterations, we further mitigated the risk of missing information in our analysis and oversimplifying the results.
The use of saturation in our analysis made sure that we did not prematurely stop including more entries and that the categories of factors were stable.

%
%

\section{Conclusions}
\label{section:conclusion}

We discussed the salient factors that are responsible for diverging results in research on TDD, and hinder the applicability of TDD research for practitioners.
These factors, extracted from literature, concern TDD definition, participants, task, type of project, and comparison.

We found that TDD is mainly boiled down to writing tests first, and how strictly its other characteristics such as refactoring is followed is not well-explained in previous research;
studies are mostly conducted with subjects who are not proficient in TDD;
studies in brownfield projects with real-world tasks are in a minority; 
a large body of research has compared TDD against traditional development techniques;
and finally, we noticed a lack of attention to the long-term effect of TDD.

We discussed the implications of this work for researchers studying TDD, and for practitioners seeking to adopt this technique. 
We hope that this work paves the way to conduct studies that produce more converging results in this field.
%

%
%

\section*{Acknowledgment}
The authors greatly appreciate the feedback from Prof. Oscar Nierstrasz and the anonymous reviewers.

\bibliographystyle{ACM-Reference-Format}
\bibliography{TDD-threats}

\end{document}